\documentclass[12pt]{spieman}  
\usepackage{amsmath,amsfonts,amssymb}
\usepackage{graphicx}
\usepackage{setspace}
\usepackage{tocloft}
\usepackage{lineno}

\def\arcdeg{\mbox{$^\circ$}}%

\title{The Focal Plane of the Arcus Probe X-Ray Spectrograph}

\author[a,*]{Catherine E.\ Grant}
\author[a]{Marshall W.\ Bautz}
\author[a]{Eric D.\ Miller}
\author[a]{Richard F.\ Foster}
\author[a]{Beverly LaMarr}
\author[a]{Andrew Malonis}
\author[a]{Gregory Prigozhin}
\author[a]{Benjamin Schneider}
\author[b]{Christopher Leitz}
\author[c]{Abraham D.\ Falcone}

\affil[a]{Kavli Institute for Astrophysics and Space Research, Massachusetts Institute of Technology, Cambridge, MA, USA}
\affil[b]{Lincoln Laboratory, Massachusetts Institute of Technology, Lexington, MA, USA}
\affil[c]{Department of Astronomy and Astrophysics, Pennsylvania State University, University Park, PA, USA}

\cftpagenumbersoff{figure}
\cftpagenumbersoff{table} 

\newcommand{\fevv}{$^{\rm{55}}$Fe\ }

\begin{document} 
\maketitle

\begin{abstract}
The Arcus Probe mission concept provides high-resolution soft X-ray and UV spectroscopy to reveal feedback-driven structure and evolution throughout the universe with an agile response capability ideal for probing the physics of time-dependent phenomena.  The X-ray Spectrograph (XRS) utilizes two nearly identical CCD focal planes to detect and record X-ray photons from the dispersed spectra and zero-order of the critical angle transmission gratings. In this paper we describe the Arcus focal plane instrument and the CCDs, including laboratory performance results, which meet observatory requirements.
\end{abstract}


{\noindent \footnotesize\textbf{*}Catherine Grant,  \linkable{cgrant@mit.edu} }

\begin{spacing}{1}   

\section{Introduction}
\label{sect:intro}
The Arcus Probe mission concept, submitted to the NASA Astrophysics Probe Explorer proposal call with expected launch in the early 2030s, will explore the formation and evolution of clusters, galaxies, and stars through high resolution X-ray and UV spectroscopy.  Arcus can reveal feedback-driven structure and evolution throughout the universe due to advances in optics and high-resolution gratings which enable breakthrough science.\cite{ArcusProbe}  The Arcus mission includes two co-aligned instruments working simultaneously to provide high resolution spectra in both the soft X-ray (12\AA--50\AA; 0.25--1~keV; R $>$ 2500) and far UV (1020\AA--1540\AA; R $>$ 17,000), with an agile response capability to address time-domain science (response to target of opportunity triggers in as little as 4 hours).  

\begin{figure}[p]
\begin{center}
\includegraphics[width=6.5in]{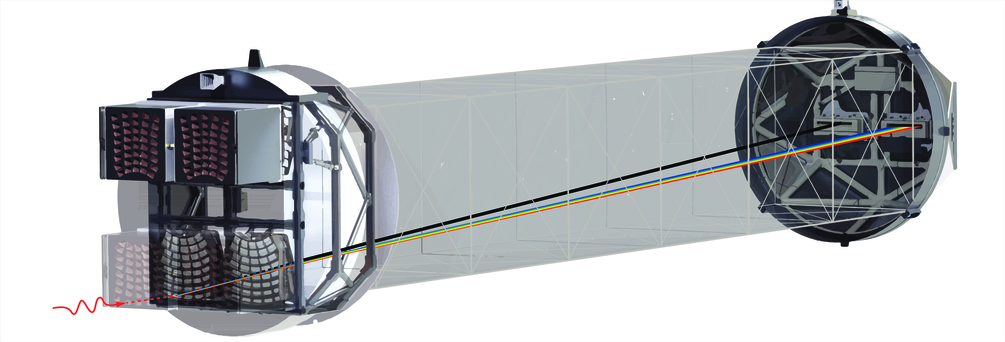}
\end{center}
\caption{A rendering of the Arcus XRS after deployment. At left the front assembly holds the four optical channels, two of which are shown with their thermal pre-colimators (left, top) and open doors, while the other two (left, bottom) have these removed to show the silicon pore optics mounted in their petal structure. Each petal holds 40 mirror modules in 8 rows, with an (unseen) grating petal mounted behind it. The extendable boom connects the front assembly to the rear assembly that holds two CCD focal planes and their associated electronics. The boom is wrapped with a light-tight sock to suppress stray light. (from Ref.~\citenum{ArcusProbe})}
\label{fig:arcusXRS}
\end{figure}

The Arcus X-ray Spectrograph (XRS) consists of four parallel, almost identical optical channels that feed two detector focal plane arrays, which record both the dispersed spectra and the zero-order of the gratings.  Each optical channel consists of silicon pore optics\cite{AthenaSPO}, identical to the design of the NewAthena optics, and critical-angle transmission gratings\cite{ArcusCAT}.  The optics plus gratings are separated from the detectors by a 10~m long coilable boom, enclosed within a sock to suppress stray light.  The rear of the spacecraft holds two CCD detector assemblies which detect and record the diffracted X-ray photons.  For a given optical channel, the zero order image and low diffraction orders fall on one detector array, while the high orders fall on the other -- two channels on one side and two channels in the reverse direction.  The optical channels are offset from each other in the cross-dispersion direction, so as not to overlap on the detector.  The CCD energy resolution is used to differentiate the co-spatial diffraction orders. A rendering of the XRS after deployment is shown in Figure~\ref{fig:arcusXRS}.

Section~\ref{sect:xintro} introduces the XRS focal plane instrument, while section~\ref{sect:ccd} further details the CCDs.  Ground testing of the CCDs designed for Arcus XRS is presented in \ref{sect:ccid94}. The Arcus mission\cite{ArcusProbe} and the UV Spectrograph\cite{ArcusUVS1,ArcusUVS2} are further described elsewhere and in this special section of JATIS.

\section{XRS Instrument Detector Subsystem}
\label{sect:xintro}

The XRS Instrument Detector Subsystem (IDS) contains the detector array CCDs and their associated electronics and structures which detect the X-ray photons focused and dispersed by the optics and gratings. The IDS consists of two Detector Assemblies (DA), each with eight CCDs, and two associated Detector Electronics (DE) boxes.  Each DA+DE pair operates independently and in parallel.  The IDS records diffracted and zero-order photons and sends the digitized pixel pulseheight data to the Event Recognition Processor (ERP) in the XRS Instrument Control Unit (XICU) where the raw data is processed and sent to the spacecraft for downlink to the ground. Some basic characteristics of the XRS focal plane are shown in Table~\ref{tab:FPchar}. 

\begin{table}[t]
\caption{XRS focal plane characteristics \label{tab:FPchar}}
\begin{center}       
\begin{tabular}{|l|l|}
\hline\hline
Detectors &Frame transfer X-ray CCDs \\ \hline
Focal plane array &8-CCD array per Detector Assembly\\ \hline
CCD spectral resolution requirement &FWHM $<$ 70 eV @ 0.5 keV\\\hline
System read noise requirement & $\leq$ 4 e- RMS @ 625 kpixels / sec\\\hline
CCD frame rate & 1 Hz\\ \hline
Focal plane temperature     & $-90 \pm 0.5$\arcdeg C \\ \hline
Optical blocking filter &40 nm Al on-chip\\\hline
Contamination blocking filter &45 nm polyimide $+$ 30 nm Al\\ 
\hline
\end{tabular}
\end{center}
\end{table} 

\subsection{Detector Assembly}

\begin{figure}[p]
\begin{center}
\includegraphics[height=3in]{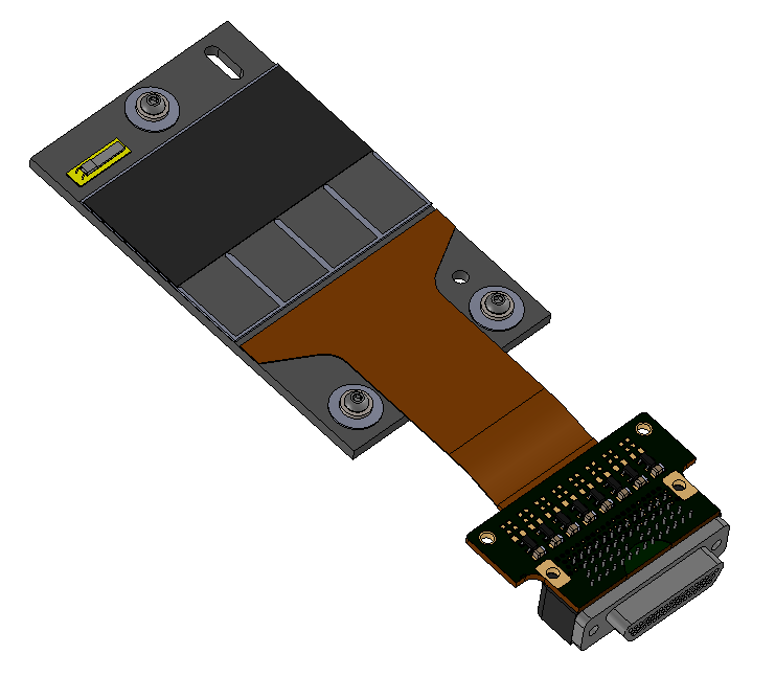}
\end{center}
\caption{Arcus MIT Lincoln Laboratory CCID-94 back-illuminated CCD in a flight package. The dark rectangle is the imaging area of the CCD; the grey smaller rectangles are the frame store areas. Each Detector Assembly has eight such CCDs.}
\label{fig:CADCCD}
\end{figure}

\begin{figure}[p]
\begin{center}
\includegraphics[width=5.5in]{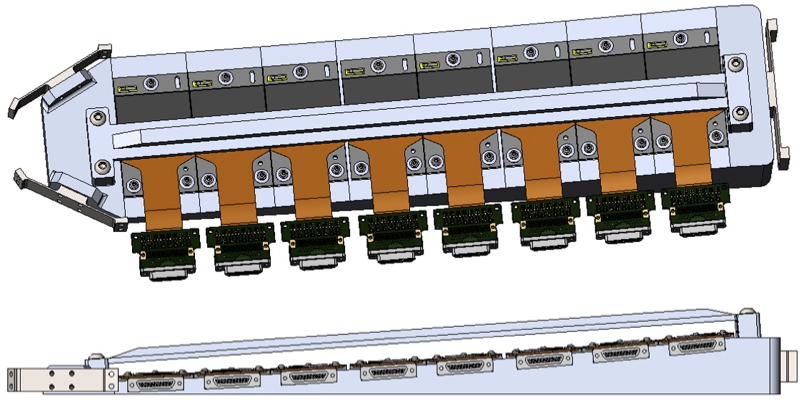}
\end{center}
\caption{Eight Arcus CCDs in a Detector Assembly focal plane arranged linearly along the dispersion direction of the gratings.  (top) Viewed from above, as seen by the optics, and (bottom) viewed from the side.  The frame store region is shielded by an aluminum cover. Each CCD is tilted to best match the curved focal surface.  Arcus contains two Detector Assemblies to read out the four optical channels.}
\label{fig:CADfocalplane}
\end{figure}

Renderings of an individual Arcus CCD in the flight package and an XRS Detector Assembly focal plane are shown in Figures~\ref{fig:CADCCD} and \ref{fig:CADfocalplane}.  Eight CCDs are arranged in a linear array along the dispersion direction of the gratings.  The CCDs are tilted to best match the curved focal surface. Gaps between the CCDs are minimized ($\sim$2~mm between active areas) and the alignment with the XRS optics has been optimized to ensure that of the four optical channels, no two spectra have chip gaps at the same wavelength. An aluminum cover shields the frame store region from X-ray interactions during readout.

\begin{figure}[p]
\begin{center}
\includegraphics[width=5.5in]{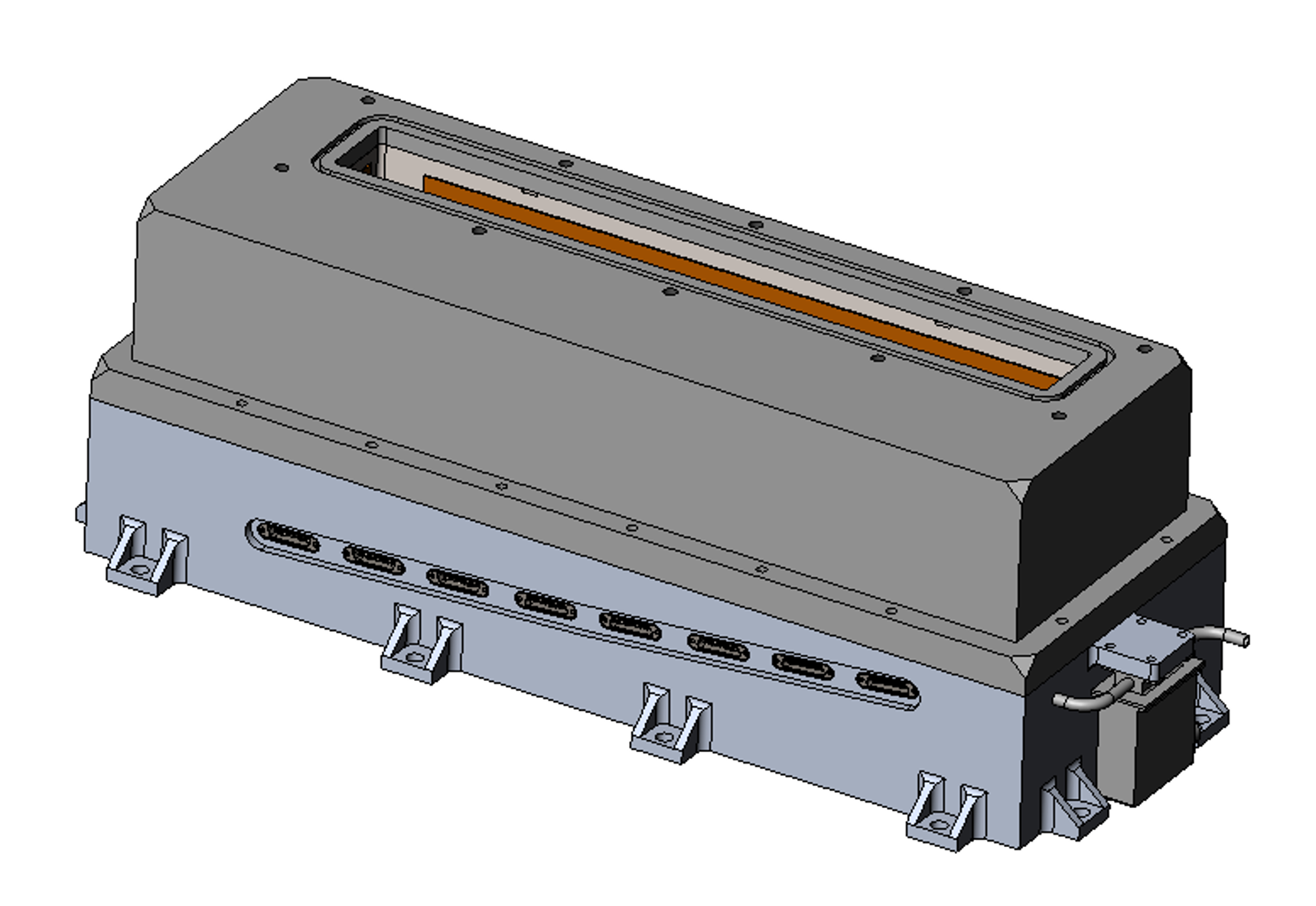}
\end{center}
\caption{One of two Arcus Detector Assemblies with the focal plane CCDs enclosed by the housing. X-rays enter through the long slot at the top of the housing and pass through the Contamination Blocking Filter, just inside. The door, which maintains a vacuum within the housing and is opened once after launch, is not shown.}
\label{fig:CADhousing}
\end{figure}

The Detector Assembly focal planes are each enclosed in a housing which provides structural support for the CCDs, thermal and electrical connections, and a controlled environment.  The detector housing is shown in Figure~\ref{fig:CADhousing}. The aluminum housing provides 5 gm~cm$^{-2}$ of (spherically averaged) radiation shielding, which meets the Arcus lifetime requirement for mitigation of radiation damage and minimization of charge transfer inefficiency (CTI) increase. An analysis of the effect of shielding on in-band particle background will be done as part of a future study. The CCDs are passively cooled via a radiator with active heater control to maintain $-90 \pm 0.5$\arcdeg C.  A door preserves the vacuum within the housing to protect the focal plane from molecular contamination and is opened once post-launch after sufficient time for the rest of the spacecraft to outgas.

Contamination Blocking Filters (CBF) are mounted inside the housing to protect the cold detectors from accumulating molecular contamination which would reduce the XRS effective area by absorbing low energy X-rays.  The CBF also provides optical light blocking. The temperature of the CBF is maintained above +20\arcdeg C by heater control to prevent contamination of the filters.  The CBF consists of a metal mesh supporting 45~nm polyimide and 30~nm aluminum and is in vacuum at launch to prevent any damage due to acoustic loading.

Optical stray light on the CCD focal plane would degrade the X-ray photon energy measurement required for order sorting by producing additional time- and position-varying background noise. A series of design elements mitigates stray light.  The XRS boom is surrounded by a 25-$\mu$m thick sock which limits stray light into the optical bench.  A baffle limits the view of each DA to the optics.  Finally, the CBF and on-chip Optical Blocking Filter (OBF) attenuate any remaining visible/IR stray light to 10 ppm of the incoming flux.

The CCD response (gain, FWHM, and QE) will be calibrated at MIT before integration into the full XRS using protocols developed for Chandra ACIS and Suzaku XIS, with X-ray line sources from 0.5 to 6~keV\cite{ACIScal,ACISBESSY,XIScal}. Pre-launch end-to-end X-ray testing of the XRS will verify system performance.  Changes in the CCD gain after launch will be tracked by collimated \fevv sources in each DA housing which produce X-rays at 5.9 and 6.4~keV from Mn-K$\alpha$ and K$\beta$. These are mounted to continuously illuminate the top portion of all CCDs outside of the region where the dispersed spectra are detected. This scheme was successfully used on Suzaku to monitor changes in the CCD performance, measure CTI, and calibrate gain. 

\subsection{Detector Electronics}

Each Detector Assembly has a corresponding Detector Electronics (DE) box that drives and reads out the CCDs, transmitting the digitized pixel data on to the XRS Instrument Control Unit (XICU).  Each DE contains eight identical Focal Plane Electronics Boards (FPEBs), one per CCD, along with an interface board, thermal control board, and power distribution board.

The FPEBs provide the programmable analog clock waveforms and biases to operate the CCDs, as well as low-noise analog signal processing and digitization of the output signal. The FPEB clock drivers consist of CMOS-level translators followed by discrete current amplifiers. A programmable sequencer provides CCD clock and analog processor timing. The low-voltage analog design reduces power and distortion, and allows lower impedances for low thermal Johnson noise and cross-talk. The CCD readout sequence and clock levels are configurable on-orbit. The circuitry has heritage from Chandra ACIS and Suzaku XIS, with a low total readout noise of $<$~4~e- which minimizes the noise contribution to the CCD energy resolution, required to separate the spatially coincident diffraction orders.  Low noise also ensures detection of X-ray events at the low-energy end of the Arcus passband (0.25~keV).  Each board has a dedicated analog processing chain serving the eight independent CCD outputs per chip. The one second frame time ensures negligible pile up in the dispersed spectra.

The DE interface board provides a command and telemetry interface between the XICU and the FPEBs. The interface board produces a stream of digitized pixel data from the CCDs and FPEBs, which is sent on to the XICU for event finding and filtering, as well as sending science and engineering housekeeping information. Regulated power is provided to the FPEBs by the power distribution board while the thermal control board regulates the temperature of the focal plane and the CBF to -90\arcdeg C and +20\arcdeg C, respectively.

\subsection{XRS Instrument Control Unit}
\label{sec:icu}

The activities of the XRS are controlled by the XRS Instrument Control Unit (XICU), which is responsible for gathering and storing all XRS data for transfer to the spacecraft.  A key component of the XICU is the Event Recognition Processor (ERP), which reads in the digitized pixel pulseheights from the CCDs and extracts candidate X-ray events, reducing the data rate by many orders of magnitude.\cite{BurrowsERP}  The ERP uses the same event recognition algorithms employed on Chandra, Swift, and Suzaku, bias-correcting the pixel stream passed from the DE, then identifying local maxima as candidate events and storing the position, time, and pulseheight values for a $3\times3$ pixel event island, for packaging into telemetry. The ERP implements event processing in firmware rather than software, achieving much higher data processing speeds than legacy missions, required for the large 32~Megapixel Arcus focal plane. The design and implementation of a prototype FPGA-based ERP are described in Ref.~\citenum{BurrowsERP} and the measured performance of prototype is sufficient to meet the needs of the Arcus XRS focal plane.

\section{XRS CCDs}
\label{sect:ccd}

The Arcus CCDs are CCID-94 backside-illuminated (BI) frame-transfer devices designed and manufactured specifically for Arcus by MIT Lincoln Laboratory (MIT/LL).  Some of their characteristics are listed in Table~\ref{tab:CCDchar}. They derive directly from the CCID-41 devices, which performed as designed throughout the 10-year Suzaku mission, and the CCID-17 devices, which are still operating on Chandra more than 24 years after launch. X-ray photons interact in the depleted silicon, producing photoelectrons that drift into the buried channel under the influence of an applied electric field.  After the one second integration time, the charge packets are quickly transferred into the frame store region and then more slowly transferred through the serial register to one of 8 output ports to be amplified and read out.

\begin{table}[t]
\caption{XRS CCD characteristics \label{tab:CCDchar}}
\begin{center}       
\begin{tabular}{|l|l|}
\hline\hline
Detectors &Back-illuminated frame transfer CCDs\\ \hline
Format &2048 $\times$ 1024 pixel imaging array \\ \hline
Image area pixel size &24 $\times$ 24 $\mu$m \\\hline
Die size &5 cm $\times$ 4 cm \\\hline
Output ports & 8\: 2-stage pJFETs \\\hline
Transfer gate design & Triple layer polysilicon \\\hline
Radiation tolerance features &Trough, charge injection \\\hline
Detector thickness &50 $\mu$m (fully depleted) \\\hline
Back surface passivation &Molecular Beam Epitaxy 10 nm \\\hline
Typical serial rate &0.5 MHz \\\hline
Typical parallel rate &0.1 MHz \\\hline
Full frame rate & 1 Hz\\ \hline
\hline
\end{tabular}
\end{center}
\end{table} 

The CCID-94 shares many design features of its predecessor BI detectors: 24~$\mu$m pixel size, 50~$\mu$m thickness (fully depleted), three-phase charge transfer through triple layer polysilicon transfer gates, three-side abuttability, and frame store design. The CCID-94 is capable of charge injection, which has been very effective at mitigating charge transfer inefficiency due to radiation damage on Suzaku XIS.\cite{BautzCI,NakajimaCI}  The Arcus devices are twice as wide as the Chandra and Suzaku devices, 2048 pixels compared to 1024, to better cover the grating dispersion pattern, and have twice as many parallel output amplifiers to maintain sufficiently fast readout speed for preventing photon pile-up while ensuring low readout noise.  

The single stage on-chip MOSFET used in Chandra and Suzaku has been replaced by a high-bandwidth, high-responsivity two-stage pJFET amplifier.  This allows up to ten times faster readout speeds while maintaining low readout noise. Arcus requires a 1~second CCD frame time to minimize pileup, which is easily met by the baseline 625~kHz serial readout rate. The CCD backside is passivated with a molecular beam epitaxy process that deposits a 10~nm layer of heavily doped silicon. The CCDs also feature a 40~nm directly-deposited aluminum blocking filter, which works in conjunction with the CBF to prevent stray optical light from producing additional noise in the detectors.\cite{BautzOBF}

\section{CCD X-ray Performance}
\label{sect:ccid94}

\begin{figure}[p]
\begin{center}
\includegraphics[width=4.5in]{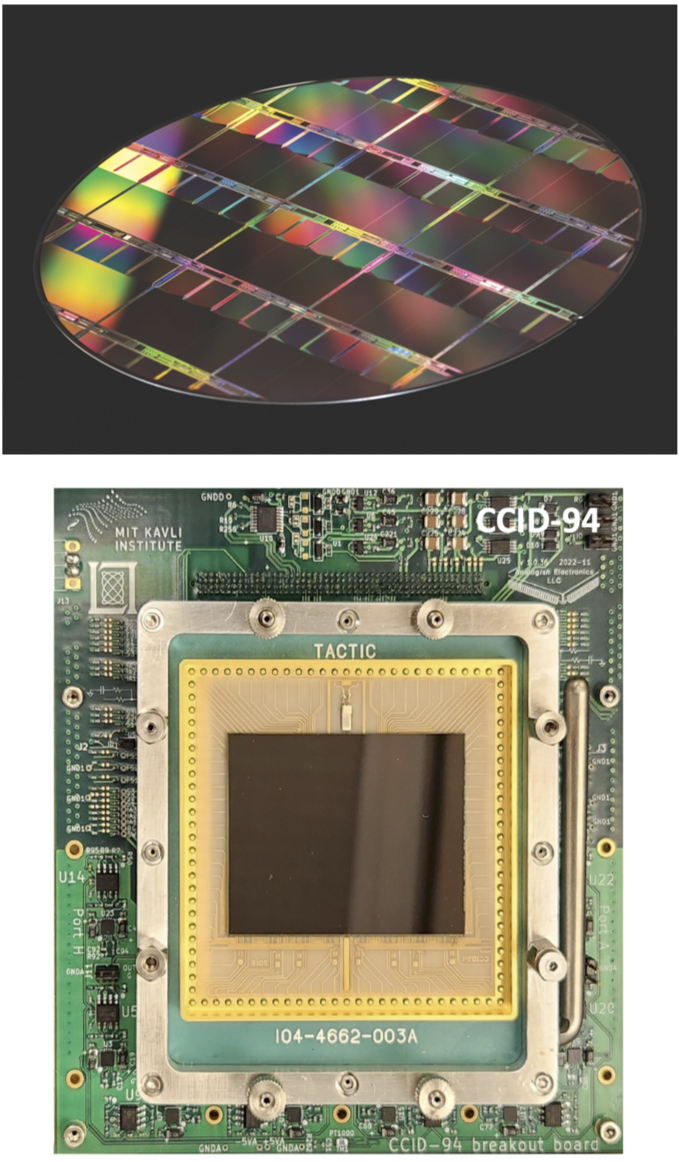}
\end{center}
\caption{(top) Photograph of MIT/LL CCID-94 wafer before packaging.  The large rectangular imaging area and smaller rectangular frame store arrays of each CCD are visible as the wafer is viewed from the front illuminated side. (bottom) Photograph of a packaged MIT/LL CCID-94 undergoing testing for Arcus development and performance demonstration at MKI. The frame store cover is not installed. As the device is back illuminated, the structures defining the imaging and frame store arrays cannot be seen.  The eight readout chains are visible at the bottom of the device.}
\label{fig:photo94}
\end{figure}

The development of advanced, fast-readout CCD detectors for future missions has been undertaken by our groups at MIT/LL and the MIT Kavli Institute (MKI) for the past several years, most recently reported in Ref.~\citenum{BautzCCD} and~\citenum{AXISCCD}.  Photographs of the CCID-94 device, designed specifically for the Arcus XRS, are shown at wafer level (front-illuminated) and after packaging for lab testing (back-illuminated) in Figure~\ref{fig:photo94}. The CCDs tested in the lab do not have the on-chip optical blocking layer that is planned for the flight devices.

The MKI facility used for X-ray performance testing of the CCID-94 is described more fully in Ref.~\citenum{AXISCCD}. Test CCDs in a vacuum chamber with a liquid-nitrogen cryostat can be illuminated by a variety of X-ray line sources, including radioactive \fevv producing Mn-K$\alpha$ and K$\beta$ fluorescence lines at 5.9 and 6.4~keV, and a radioactive $^{210}$Po source fluorescing a Teflon target producing lines of C-K (0.25~keV) and F-K (0.7~keV). In addition, there is an In-Focus Monochromator (IFM) which produces clean monochromatic lines at energies below 2~keV.\cite{IFM}

\begin{figure}[t]
\begin{center}
\includegraphics[width=.95\linewidth]{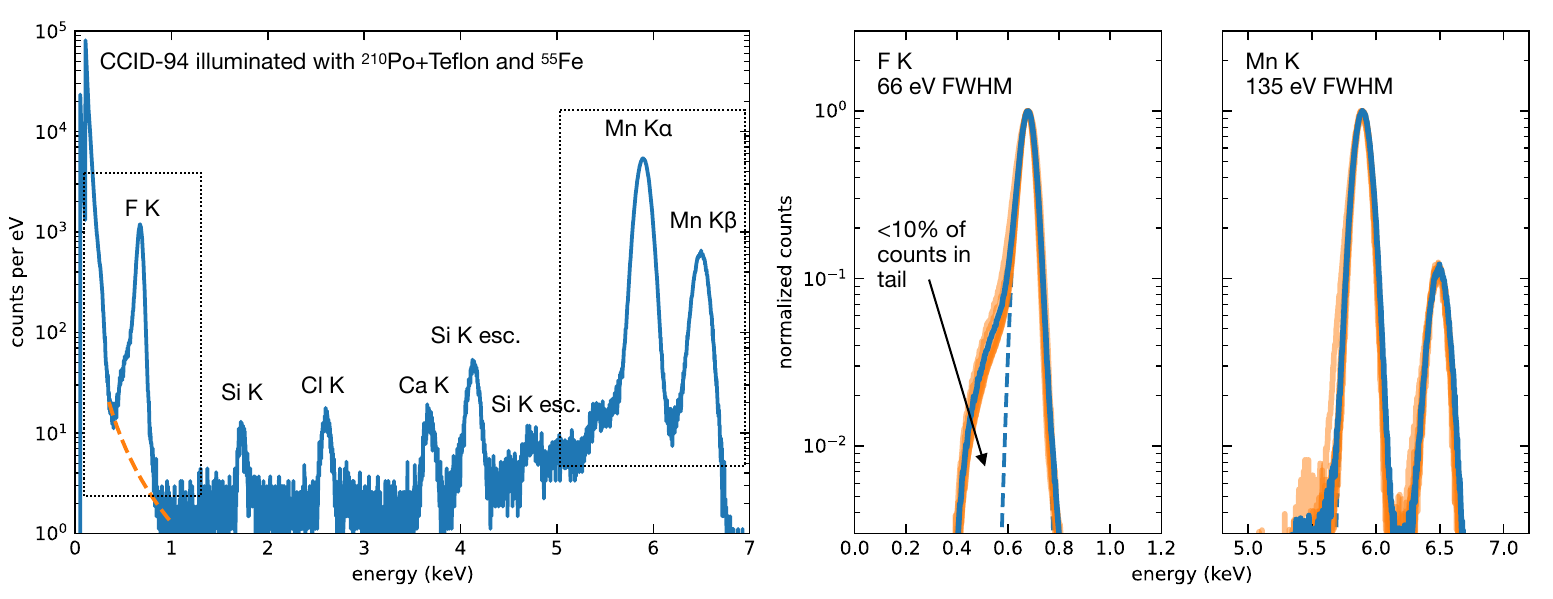}
\end{center}
\caption{(left panel) Spectrum of a single CCID-94 segment (node `C') simultaneously illuminated by $^{210}$Po with a Teflon target and $^{55}$Fe. The primary fluorescence lines of F-K, Mn-K$\alpha$ and Mn-K$\beta$ can be seen along with several other fluorescence and escape features. (right two panels) Zoom-in of the F-K and Mn-K peaks from the left panel, now also showing spectra from the other seven nodes of this device in orange. The F-K line has been corrected for the noise continuum by subtracting a best-fit power law, shown in dashed orange in the left panel. Gaussian fits (dashed blue lines) indicate that the FWHM meets the Arcus spectral resolution requirement. While the F-K peak shows a non-Gaussian tail, it contains fewer than 10\% of the line counts. These results include events with pixel multiplicities up to four.}
\label{fig:94pofe55}
\end{figure}

The measured readout noise is low and uniform across the device, ranging from 2-3 e- RMS, well below the 4 e- readout noise requirement. The spectral resolution is also excellent across the Arcus energy bandwidth. Figure~\ref{fig:94pofe55} demonstrates CCID-94 FWHM performance using a combination of the two radioactive sources. The serial and parallel clocking rates and the CCD temperature were all consistent with those planned for Arcus. The data have been processed in a similar fashion as would be done on-board: finding candidate X-ray events, recording a $3 \times 3$ pixel event island, and calculating the event energies. Each event is assigned a pixel ``multiplicity", akin to the ``grade" on ASCA, Chandra, and Suzaku, to indicate the number of pixels in an event island that are above threshold. We include multiplicities up to four in our results. The measured FWHM at 0.7~keV, 66~eV, meets the Arcus low energy spectral resolution requirement of 70~eV with similar performance on all eight nodes of the device. A small non-Gaussian tail develops at energies below 1~keV, which may be due to incomplete charge collection due to losses at the surface of the CCD but represents less than 10\% of the photon counts in the line at 0.7~keV. This feature is currently under further investigation which will  be reported in a future paper.

\begin{figure}[t]
\begin{center}
\includegraphics[height=4in]{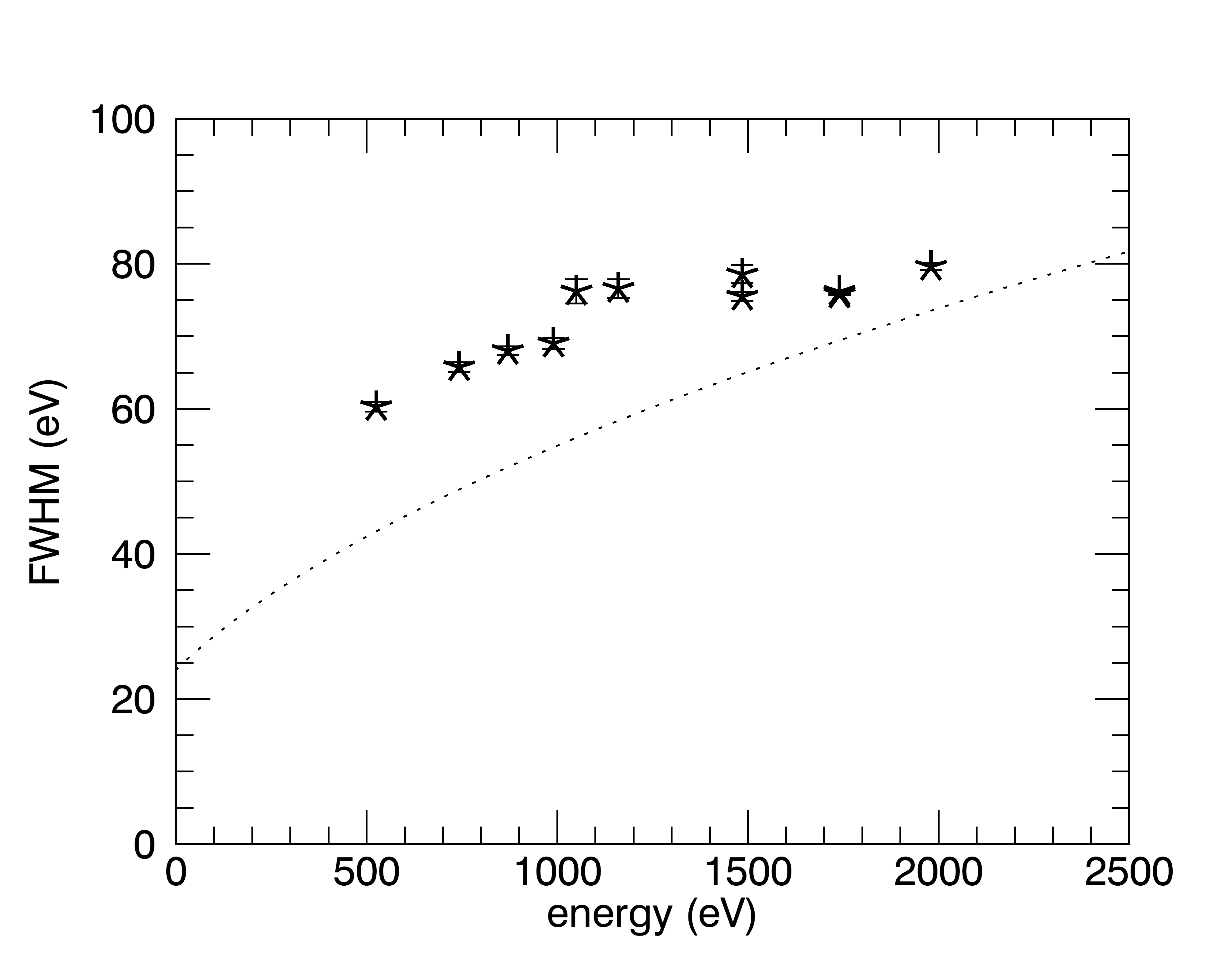}
\end{center}
\caption{Spectral FWHM for the CCID-94 as a function of X-ray line energy as measured by the IFM.  The dotted line is the theoretical best possible FWHM assuming the measured noise for this output node (2 e-) and a simplified model of event splitting. The Arcus requirement of FWHM $<$70~eV at 0.5~keV is clearly met by this device. These results include events with pixel multiplicities up to four as is typical for Chandra and Suzaku.}
\label{fig:94ifm}
\end{figure}

A second demonstration of the CCID-94 performance was done with the IFM, measuring the FWHM at several X-ray energies from 0.5 to 2~keV.  This is shown in Figure~\ref{fig:94ifm}, where the spectral FWHM remains below the Arcus requirement up to $\sim$~1~keV.

\section{Summary}

The Arcus Probe mission concept will provide high resolution X-ray and UV spectroscopy to explore a wide range of science topics.  The X-ray Spectrograph provides substantial improvement in sensitivity and resolution over anything flown previously.  The XRS focal plane utilizes high heritage MIT/LL CCDs with proven technologies to read out the diffracted photons.  Lab testing confirms CCID-94 performance more than meets the required CCD spectral resolution and read out noise for low-energy sensitivity and order sorting of the grating photons.

\subsection*{Disclosures}
The authors of this paper are members of the Arcus collaboration. Should NASA select Arcus for implementation, their institutions will receive funding which may be used to fund the author's salaries in full or in part in the future.

\subsection* {Code and Data Availability} 
The data that support the findings of this article are not publicly available. They can be requested from the author at cgrant@mit.edu.

\subsection* {Acknowledgments}
We gratefully acknowledge support from NASA through the Strategic Astrophysics Technology (SAT) program, grants 80NSSC19K0401 and 80NSSC22K0788, and from the Kavli Research Infrastructure Fund of the MIT Kavli Institute for Astrophysics and Space Research. We thank the reviewers for helpful comments.


\bibliography{report}   
\bibliographystyle{spiejour}   





\end{spacing}
\end{document}